\documentclass[twocolumn,showpacs,preprintnumbers,amsmath,amssymb,A4paper]{revtex4}

\usepackage{graphicx}
\usepackage{dcolumn}
\usepackage{bm}

\begin{document}
\newcommand{\kp}{{\bf k$\cdot$p}\ }
\newcommand{\Pp}{{\bf P$\cdot$p}\ }

\preprint{APS/123-QED}
\title{Oblique spin-flip reflection of relativistic electrons from a potential barrier}

\author{Wlodek Zawadzki and Pawel Pfeffer}
 \affiliation{Institute of Physics, Polish Academy of Sciences\\
 Al.Lotnikow 32/46, 02--668 Warsaw, Poland\footnotetext{$^*$ e-mail address: zawad@ifpan.edu.pl}\\}
\date{\today}

\begin{abstract}

The Dirac equation is used to describe oblique spin-conserving and
spin-flip reflections of relativistic electrons from a one-dimensional
potential barrier in a vacuum. When an electron hits the barrier from an
oblique direction, its effective spin-up and spin-down states have different
energies due to the spin-orbit interaction and the fact that the system
has a structural inversion asymmetry. When the electron is reflected in an
elastic spin-conserving process, incoming and outgoing angles are the
same. If an elastic reflection is a spin-flip process, the outgoing angle
is different from the incoming one. As a result, a beam of incoming
spin-polarized electrons is split after the reflection into two beams.
Angles and intensities of the reflected beams are calculated. It is shown that the reflection can be used to polarize and filter effective electron spins. Spin
splitting of energies for relativistic electrons due to the spin-orbit
interaction in an asymmetric quantum well is described.

\end{abstract}

\maketitle
\ \\
\section{\label{sec:level1}INTRODUCTION\protect\\ \lowercase{}}
\ \\

The Dirac equation (DE) is the basic tool for description of relativistic
electrons in a vacuum in the presence of external fields. Its solutions
for electrons in the presence of uniform electric and magnetic fields, the
Coulomb potential, the Kronig-Penney periodic potential etc. are known,
see e.g. [1, 2]. The Dirac equation describes the electron
spin $s = 1/2$ so that, if an electron moves in an electric potential, one
deals with the spin-orbit interaction. In the so called weakly
relativistic limit of DE, the spin-orbit term appears explicitly among
three terms of the standard  $v^2/c^2$ expansion [3]. However, one
should bear in mind that the spin-orbit interaction in DE appears
automatically once we deal with an electric potential. A good example is
given by DE for the electron in the Coulomb potential. Energies and
solutions for this case are known exactly and they contain the well known
effects of spin-orbit interaction (SOI). Still, the SOI has a specific
symmetry and for some system geometries it gives no effects.  For example,
in the often considered reflection of relativistic electrons from a
barrier, the electrons come and are reflected along the direction
perpendicular to the barrier and  the effects of SOI are not felt.
However, if the electrons come to the barrier from an oblique direction,
the effects of SOI appear and the energy spectrum is split for two spin orientations. This property of DE was overlooked in the work
of Glass and Mendlowitz [4] and it is our purpose to correct this description.
\ \\

Our main point can be briefly illustrated by considering the spin-orbit interaction $H^{so}$, resulting from the Dirac equation in the presence of an
external one-dimensional potential $V(z)$. In this case the SOI takes the form
\ \\
\begin{equation}
\hat{H}_{so} = \frac{\hbar}{4m_0^2c^2} \frac{\partial V(z)}{\partial z}(\hat{p}_x \sigma_y - \hat{p}_y \sigma_x)
\;\;,
\end{equation}
\ \\
in the standard notation. For $V(\textbf{r}) = V(z)$ there is $p_x \rightarrow \hbar k_x$ and $p_y \rightarrow \hbar k_y$, so that, if $k_x$ or $k_y$ components do not vanish, i.e. the incidence is oblique, the spin-orbit energy in general does not vanish. In this case elastic spin-conserving and spin-flip electron reflections have different outcoming directions.

In addition, we want to consider more generally effects of the SOI for electrons
in systems with a spatial inversion asymmetry (SIA). Such systems play
an important role in the modern physics of solids, see the review [5], and they could
become important also for systems involving relativistic electrons.

Our paper is organized as follows. In Section II we describe oblique spin-conserving and spin-flip electron reflections from the barrier according to the Dirac equation. The spin-orbit energy is estimated in Section III for the case when the unperturbed electron spectrum is continuous and a numerical example is given for the reflection. In Section IV we describe the spin splitting of electron energy due to the SOI in an asymmetric quantum well. We discuss our results in Section V. The paper is concluded by a summary. In Appendix we consider an almost nonrelativistic approximation to the reflection from the barrier.
\ \\
\section{\label{sec:level1}SPIN-CONSERVING AND SPIN-FLIP REFLECTION\protect\\ \lowercase{}}
\ \\

We describe oblique spin-conserving and spin-flip reflections from a barrier using the Dirac equation.
It is well known that stationary DE, which is originally a set of four first-order differential equations, can be separated into two sets of two equations each, corresponding to positive and negative electron energies. One can then look for solutions by substitution and obtain one set of two second-order equations for positive electron energies of the following form, see [6]
\ \\
\begin{equation}
\left [
({\bm{\hat{\sigma}}} \cdot {\bf{\hat{p}}})\frac{1}{2m(V)}({\bm{\hat{\sigma}}} \cdot {\bf{\hat{p}}})+V(\bf{r})
\right ] \Phi=E \Phi
\;\;,
\end{equation}
\ \\
where $m(V) = m_0\{1 + [(E - V(\textbf{r})]/2m_0c^2\}$ and $\Phi$ is a two-component spinor. The energy $E$ does not contain the rest electron energy. Equation (2) contains all the information of the Dirac equation. In fact, it may also give additional spurious solutions, see [7].

We consider a one-dimensional case $V(\textbf{r}) = V(z)$, in which one can separate $x$ and $y$ variables, so that $p_x \rightarrow \hbar k_x$ and $p_y \rightarrow \hbar k_y$. Then Eq. (2) can be rewritten in the form
\ \\
$$
\left\{\left[ \frac{-\hbar^2}{2}\frac{\partial}{\partial z}\frac{1}{m(z)}\frac{\partial}{\partial z} + V(z) - E'\right]\left[\begin{array}{cc}
1&0 \\0&1\\ \end {array}\right] +\right.
$$
\ \\
\begin{equation}
\left.-\frac{\hbar^2}{2}
\left[\frac{\partial}{\partial z}\frac{1}{m(z)}\right]
\left[\begin{array}{cc} 0 &ik_-\\-ik_+& 0 \\ \end {array}\right]\right\}\left|\begin{array}{c} a\Phi_{\uparrow} \\ b\Phi_{\downarrow}\end{array}\right| = 0
\;\;,
\end{equation}
\ \\
where $k_{\pm} = k_x \pm i k_y$, $E' = E - \hbar^2(k_x^2 + k_y^2)/2m(z)$, $a$ and $b$ are coefficients to
be determined. Further
\ \\
\begin{equation}
m(z) = m_0\left[1+\frac{E-V(z)}{2m_0c^2}\right]\;\;\;.
\end{equation}
\ \\
The differentiation $\partial/\partial z$ in the second term of Eq. (3) acts only on $1/m(z)$. In the nonrelativistic limit: $E - V \ll 2m_0c^2$, there is $m(z) \approx m_0$. In the second (spin-orbit) term of Eq. (3), one performs first the differentiation $\partial/\partial z [1/m(z)]$. If one deals with the barrier, for which $V(z) = 0$ for $z \le 0$ and $V(z) = V_b$ for $z > 0$, we have free electron solutions for $z \le 0$ and decaying exponential solutions for $z > 0$.

We deal in Eq. (3) with two identical diagonal terms $\hat H_0$ and the nondiagonal terms $\hat H_{so}$ corresponding to the spin-orbit interaction (cf. Eq. (A.1) in Appendix). In the absence of $\hat H_{so}$ the two unperturbed spin states
\ \\
\begin{equation}
\Phi_{\uparrow} = \Phi\left(\begin{array}{c} 1 \\ 0\end{array}\right)\;\;\;\;\;\;\;\Phi_{\downarrow} = \Phi\left(\begin{array}{c} 0 \\ 1\end{array}\right)
\;\;,
\end{equation}
are degenerate and have the energy $E_0$. If $\hat H_{so}$ does not vanish we follow the standard procedure for the perturbation of two degenerate states, see for example Ref. 8 or Ref. 9. According to this scheme the perturbed energies are
\ \\
\begin{equation}
E_{1, 2} = E_0 \pm\left|\hat H^{\uparrow\downarrow}_{so}\right|\;\;,
\end{equation}
where $E_0 = H^{\uparrow\uparrow}_0 = H^{\downarrow\downarrow}_0$ is the matrix element of the diagonal term taken between $\Phi_{\uparrow}$ or $\Phi_{\downarrow}$ functions. Thus the splitting of $E_0$ is given by $2\Delta = 2\left|\hat H^{\uparrow\downarrow}_{so}\right|$, i. e. by the double matrix element of the nondiagonal perturbation taken between the orthogonal unperturbed functions $\Phi_{\downarrow}$ and $\Phi_{\uparrow}$. The perturbed wave functions corresponding to the energies $E_{1, 2}$ are
\begin{equation}
\Phi^{1, 2} = (\Phi_{\uparrow} \pm\Phi_{\downarrow})/\sqrt 2\;\;,
\end{equation}
independently of the perturbation strength. Using Eq. (5) one obtains
\begin{equation}
\Phi^1 = \Phi\frac{1}{\sqrt 2}\left(\begin{array}{c} 1 \\ 1\end{array}\right)\;\;\;\;\;\;\;\Phi^2 = \Phi\frac{1}{\sqrt 2}\left(\begin{array}{c} 1 \\ -1\end{array}\right)
\;\;.
\end{equation}
It is explicitly seen in Eq.(7) that the perturbed functions $\Phi^1$ and $\Phi^2$ are spin-mixed states. Since they correspond to two (and only two) different energies $E_1$ and $E_2$, they are often called "effective spin-up" and "effective spin-down" states, respectively. (Sometimes the term "pseudo-spin states" is used.) In the following we will use the term "effective spin states".

We consider a spin-up electron coming to the barrier from an oblique direction.
Without loss of generality we can choose the coordinate system in such a way that $k_y = 0$, while $k_x \ne 0$ and $k_z \ne 0$. We assume that the electron energy $E$ is smaller than $V_b$, so that for $z \le 0$ the value of $k_z$ is real, while for $z > 0$, when the electron penetrates the barrier, its wave vector $q_z$ is imaginary. Also, for $V_b < 2m_0c^2 + E$ the problem of Klein paradox does not come into play.

\begin{figure}
\includegraphics[scale=0.45,angle=0, bb = 700 20 202 540]{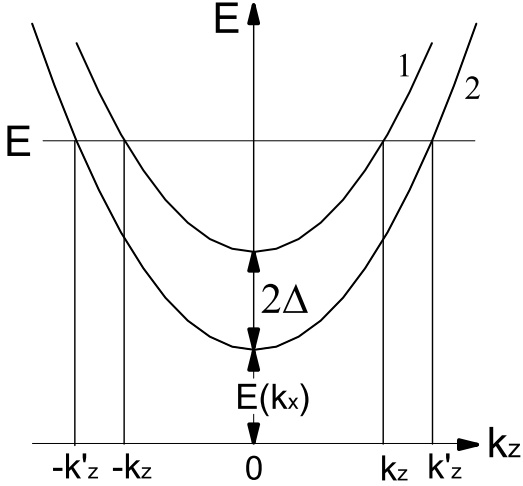}
\caption{\label{fig:epsart}{Electron energy versus wave vector $k_z$ for effective spin-up and spin-down states (schematically). The spin-orbit energy is $\Delta$, and $E(k_x) = \hbar^2k_x^2/[2m_0(1+E/2m_0c^2)]$ . For a spin-conserving elastic reflection of spin-up electron the outgoing $k_z$ is equal to $-k_z$, for a spin-flip reflection the outgoing wave vector is $-k'_z$ such, that $|k'_z| > |k_z|$. Situation for an initial spin-down electron with the initial wave vector $k'_z$ is also shown.}} \label{fig1th}
\end{figure}
The function $\Phi$ in Eq. (5) is a solution of the Schrodinger equation with the diagonal part of the Hamiltonian $H_0$ defined by Eq. (3). The effective spins are parallel to the $y$ direction.
Once $E$ and $\Delta$ are fixed, the wave vector $k_z$ can be determined from Eqs. (3) and (4). For the incoming and reflected spin-up components there is, for the left of the barrier,
\ \\
\begin{equation}
\frac{\hbar^2 k^2}{2m_0} \frac{2m_0 c^2}{2m_0 c^2+E}+\Delta = E \;\;,
\end{equation}
\ \\
where $k^2 = k_x^2 + k_z^2$. This gives
\ \\
\begin{equation}
k^2_z=\frac{(E - \Delta)(E + 2m_0c^2) - c^2 \hbar^2 k^2_x}{c^2 \hbar^2}\;\;\;,
\end{equation}
\ \\
while for the reflected spin-down component there is
\ \\
\begin{equation}
k'^2_z=\frac{(E + \Delta)(E + 2m_0c^2) - c^2 \hbar^2 k^2_x}{c^2 \hbar^2}\;\;\;.
\end{equation}
\ \\
For the effective spin-up and spin-down components penetrating the barrier there is, respectively
\ \\
\begin{equation}
q^2_z=\frac{(E - \Delta - V_b)(E - V_b + 2m_0c^2) - c^2 \hbar^2 k^2_x}{c^2 \hbar^2}\;\;\;,
\end{equation}
\ \\
and
\ \\
\begin{equation}
q'^2_z=\frac{(E + \Delta - V_b)(E - V_b + 2m_0c^2) - c^2 \hbar^2 k^2_x}{c^2 \hbar^2}\;\;\;.
\end{equation}
\ \\

Now we turn to the amplitudes of the functions. Let us again consider an incoming spin-up electron. On the left of the barrier, for $z\le 0$, the electron is described by the plane waves, the incoming wave vector is $k_z$, the reflected one with the same spin is $-k_z$, while the reflected one with the opposite spin is $k'_z$. The value of $k_x$ does not change. In addition, there exist two decaying waves penetrating the barrier: the spin-up wave with imaginary wave vector $q_z$ and the spin-down wave with the imaginary wave vector $q'_z$.
The complete wave functions for the incoming electron with the effective spin up and spin down are
\begin{widetext}
\begin{equation}
\Psi^1 = C e^{ik_x x}\left\{\left[e^{ik_z z}\left(\begin{array}{c}1\\1\end{array}\right) +
R e^{-ik_z z}\left(\begin{array}{c}1\\1\end{array}\right)+R' e^{-ik'_z z}\left(\begin{array}{c}1\\-1\end{array}\right)\right]_{| z\le 0}+
\left[T e^{iq_z z}\left(\begin{array}{c}1\\1\end{array}\right)+T' e^{iq'_z z}\left(\begin{array}{c}1\\-1\end{array}\right)\right]_{| z > 0}\right\}
\;\;,
\end{equation}
\ \\
\begin{equation}
\Psi^2 = C e^{ik_x x}\left\{\left[e^{ik'_z z}\left(\begin{array}{c}1\\-1\end{array}\right) +
P e^{-ik'_z z}\left(\begin{array}{c}1\\-1\end{array}\right)+P' e^{-ik_z z}\left(\begin{array}{c}1\\1\end{array}\right)\right]_{| z\le 0}+
\left[F e^{iq'_z z}\left(\begin{array}{c}1\\-1\end{array}\right)+F' e^{iq_z z}\left(\begin{array}{c}1\\1\end{array}\right)\right]_{| z > 0}\right\}
\;\;,
\end{equation}
\end{widetext}
\ \\
where the amplitudes $R$, $R'$, $T$, $T'$ and $P$, $P'$, $F$, $F'$  relate to the contributions mentioned above. We have assumed that both electron states have the same energies $E$ and $k_x$ values, see Fig. 1. In the following we consider explicitly only the incoming spin-up electron characterized by $\Psi^1$.
The amplitudes can be determined from the boundary conditions for the wave functions and their derivatives at $z = 0$.

The boundary conditions for the continuity of the wave function at $z = 0$ are
\ \\
\begin{equation}
 {\Psi^{1 \;up}}_{| z = 0^-} = {\Psi^{1 \;up}}_{| z = 0^+}\;\;,
\end{equation}
\ \\
\begin{equation}
 {\Psi^{1 \;low}}_{| z = 0^-} = {\Psi^{1 \;low}}_{| z = 0^+}\;\;\;,
\end{equation}
where $up$ and $low$ refer to the upper and lower components.
The remaining boundary conditions are obtained by integrating Eq. (3) across the interface at $z = 0$, separately for upper and lower components. This gives
\ \\
\begin{equation}
\frac{\partial}{\partial z}\Psi^{1 \;up}_{| z=0^-} = M\frac{\partial}{\partial z}\Psi^{1 \;up}_{|z=0^+}
+ iS\Psi^{1 \;low}_{|z=0} \;\;\;,
\end{equation}
\ \\
and
\ \\
\begin{equation}
\frac{\partial}{\partial z}\Psi^{1 \;low}_{|z=0^-} = M\frac{\partial}{\partial z}\Psi^{1 \;low}_{|z=0^+}
- iS\Psi^{1 \;up}_{|z=0} \;\;\;,
\end{equation}
\ \\
where $ M = m|_{0^-}/m|_{0^+} = E_E/E_V$ and $S = k_xV_b/E_V$, in which  $E_E = E + 2m_0c^2$ and $E_V = E - V_b + 2m_0c^2$.
\ \\
By using the above boundary conditions we obtain: $1 + R + R' = T + T'$ and $1 - R - R'= T - T'$, which gives $R' = T'$ and $R = T - 1$. From Eqs. (18) and (19) we have
\ \\
\begin{equation}
ik_z(1-R)-ik'_zR'=iMq_zT+iMq'_zT'+iS(T-T')\;\;,
\end{equation}
\begin{equation}
ik_z(1-R)+ik'_zR'=iMq_zT-iMq'_zT'-iS(T+T')
\;\;,
\end{equation}
and for the amplitudes defined in Eq. (14) one obtains
\ \\
\begin{equation}
R = \frac{[(k_z-Mq_z)(k'_z + Mq'_z)-S^2]}{[(k_z+Mq_z)(k'_z+Mq'_z)+S^2]}\;\;,
\end{equation}
\ \\
\begin{equation}
R' = \frac{-2S k_z}{[(k_z+Mq_z)(k'_z+Mq'_z)+S^2]}\;\;,
\end{equation}
\ \\
\begin{equation}
T  = \frac{2k_z(k'_z+Mq'_z)}{[(k_z+Mq_z)(k'_z+Mq'_z)+S^2]}
\;\;,
\end{equation}
\ \\
\begin{equation}
T' = R'\;\;.
\end{equation}
\ \\
The main point is that $R' \ne 0$, see also Ref. 4. According to Eq. (14) this means that the reflected spin-flip component exists. The spin-flip amplitude is proportional to $S = k_xV_b/E_V$, which comes from the spin-orbit interaction. However, one needs also $k_x \ne 0$. This agrees with the common knowledge that, for the incoming direction perpendicular to the barrier, there is no spin-flip reflection. The main physical result is that for incoming electrons with a specific polarization of the effective spin, there appear two beams of reflected electrons: one having the same effective spin and another having the opposite effective spin.

A similar reasoning for the initial spin-down electron state gives $P = R, P' = -R', F = T, F' = -T'$, where $P, P', F$ and $F'$ are the amplitudes of
$\Psi^2$ wave function written in Eq. (15). Thus the above conclusions are symmetric with respect to the effective electron spin.  However, as follows from our kinematic considerations illustrated in Fig. 1, for an incoming spin-down electron with the initial wave vector $k'_z$ the reflected spin-up beam is characterized by $-k_z$ with $|k_z| < |k'_z|$, contrary to the initial spin-up case.
\begin{figure}
\includegraphics[scale=0.45,angle=0, bb = 700 20 202 540]{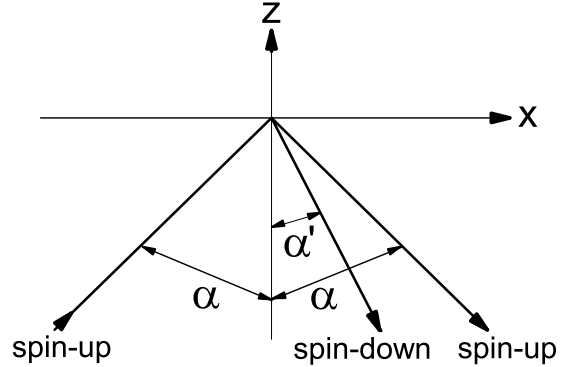}
\caption{\label{fig:epsart}{Geometry of spin-polarized relativistic electrons reflected from a
potential barrier (schematically). In a spin-conserving reflection of the spin-up electron the
outgoing angle is equal to the incoming angle $\alpha$, in a spin-flip
reflection the outgoing angle $\alpha'$ is smaller: $\alpha' < \alpha$. The effect is due to the spin-orbit interaction manifested in
oblique motion, see Fig. 1.
}} \label{fig2th}
\end{figure}

It should be clear from the above considerations that the electrons, even if they come from far way, are subjected to SOI and represent spin-mixed states. Thus, it is the effective spin that is flipped or conserved in the reflection. If the incoming electrons do not have well defined effective spin, one can consider them to be combinations of spin-up and spin-down components. For the spin-flip processes, the reflected spin-down component has the direction closer to normal (see Fig. 2), the spin-up component has the direction further from the normal, and the spin-conserving processes give the reflection having the same angle as the incoming beam. Thus the reflected beams on both sides contain spin-polarized electrons while the middle beam contains unpolarized electrons. This means that the reflection by a potential barrier can be used as a source of spin-polarized electron beams.
\ \\
\section{\label{sec:level1}Spin-orbit energy and numerical estimations\protect\\ \lowercase{}}
\ \\
\begin{figure}
\includegraphics[scale=0.45,angle=0, bb = 700 20 202 540]{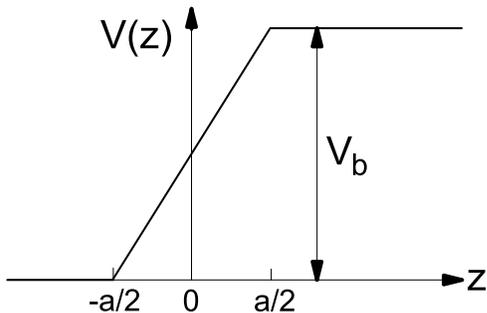}
\caption{\label{fig:epsart}{Finite one-dimensional barrier used for the estimation of spin-orbit energy (schematically). The slope is $\partial V/\partial z = V_b/a$. }} \label{fig3th}
\end{figure}\ \\
The above theory is not complete until one calculates the spin-orbit energy (SOE) in cases of interest. In principle SOE is given by the matrix element of SOI as given by Eq. (1) and, in the full relativistic description, as given by Eq. (3).
The calculation of SOE in this case is not trivial because the unperturbed spectrum of incoming and reflected electron is continuous. We do not go here into an explicit calculation but make some remarks and estimate SOE according to Eq. (1). First, it is clear that a vertical barrier corresponds to an infinite electric field since $-\partial V/\partial z = e \cal E$ is infinite. In order to avoid this infinity we consider a barrier of a finite slope linear in $z$, rising from 0 to $V_b$ over the width $a$, as illustrated in Fig. 3. Then $\partial V/\partial z = V_b/a$ for $-a/2 \ge z \le a/2$ and it vanishes outside this range. In consequence, in this range of $z$ values the SOI is
\begin{equation}
H_{so} =  \frac{\hbar^2}{4m_0^2c^2}k_x\frac{V_b}{a}
\end{equation}
\ \\
since we set $k_y = 0$, see above.
Next one should determine the eigenfunctions $\Psi(z)$ of the diagonal terms $H_o$ in Eq. (3) and calculate the $z$-part of the matrix element.
We do not carry this procedure but simply assume that $\Delta$ is given by Eq. (26). This overestimates somewhat the true value of $\Delta$ since $H_{so}$, as given by Eq. (26), does not depend on $z$, but is nonzero only in the range $-a/2 \ge z \le a/2$, while the function $\Phi$ is normalized in the whole axis $-\infty > z < \infty$. The function $\Phi$ decays quite fast to the right of the barrier, so neglecting this part gives small corrections to the normalization, while they are larger to the left of the barrier where one deals with free electron states.

Now we carry numerical estimations to give an idea about the involved orders of magnitude. Let us take $V_b = 6\cdot10^4$ eV and $a = 10^{-9}$ cm corresponding to a very high electric field $V_b/a = 6\cdot10^{13}$ eV/cm. If an incoming electron is characterized by $k_x = 10^{10}$ cm$^{-1}$ and $k_z = 5\cdot10^9$ cm$^{-1}$, the resulting SOE is $\Delta$ = 223.65 eV and the spin-up electron energy is approximately $E_1 =  \hbar^2/(k_x^2 + k_z^2)/2m_0 + \Delta \approx 4.76\cdot10^4$ eV. The incoming angle with the normal to the barrier is
$\alpha$ = ctg$^{-1} k_z/k_x = 63.43^{\circ}$ and the outgoing angle after spin-flip process is: $\alpha'$ = ctg$^{-1} k'_z/k_x = 62.90^{\circ}$, so that the difference of angles between spin-conserving and spin-flip reflected electron beams is $\alpha - \alpha' = 0.53^{\circ}$. The situation is schematically shown in Fig. 2.
As to the amplitudes of reflected beams, the corresponding quantities are: $S = 5.871\cdot10^8$ cm$^{-1}$, $|q_z| = 11.54\cdot10^9$ cm$^{-1}$,  $|q'_z| = 11.485\cdot10^9$ cm$^{-1}$, so that $R'/R = 0.037$.
\ \\
\section{\label{sec:level1} Asymmetric quantum well\protect\\ \lowercase{}}
\ \\
The previous subsections dealt with SOE for nonquantized electron states. Here we consider the case of bound electron states. This situation is again nontrivial but for a different reason. It is known that in a one-dimensional bound state, considered either classically or quantum mechanically, an average force acting on a particle vanishes. This result is intuitively quite obvious, but it can be proven rigorously, see e. g. [10]. If the only source of force is an electric field, it then follows that an average electric field in a bound state vanishes. This means that in the above situation the spin-orbit energy, which is proportional to the electric field, also vanishes. However, this result is not true in relativistic mechanics, in which the particle mass depends on velocity and potential, see Eq.  (4). It is known that in relativity there exists an additional "mass term" in the force. Since it is the total force that must vanish in a bound state, the average electric force $e \overline{\cal E}$ compensates the mass term, i. e. the average electric field does not necessarily vanish [10, 11]. Thus, in order to have a nonvanishing SOE for bound states in a quantum well one needs a relativistic approach.

We consider an asymmetric quantum well shown in Fig. 4. It has the width $a$ and the asymmetry is introduced by different heights of potential barriers on both sides. The asymmetry is necessary since, in a symmetric well, an average electric field would obviously vanish. In order to calculate the effect of SOI on the electron energy we first solve the differential equation given by the diagonal term in Eq. (3)
\begin{figure}
\includegraphics[scale=0.45,angle=0, bb = 700 20 202 540]{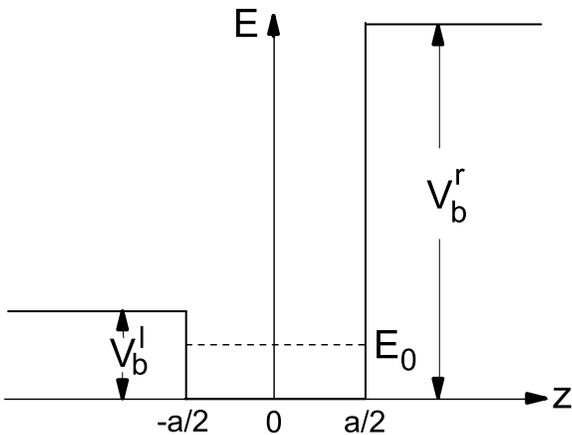}
\caption{\label{fig:epsart}{Asymmetric one-dimensional quantum well used in the calculation of
spin-orbit energy for relativistic electrons (schematically). Value of $E_0$
includes the energy of bound state and kinetic energy of the transverse
motion characterized by momentum $\hbar k_x$.
}} \label{fig4th}
\end{figure}
\ \\
\begin{equation}
\left[ \frac{-\hbar^2}{2}\frac{\partial}{\partial z}\frac{1}{m(z)}\frac{\partial}{\partial z} +\frac{\hbar^2(k_x^2 + k_y^2)}{2m(z)}
 + V(z) - E\right]\Psi(z) = 0 \;\;.
\end{equation}
 \ \\
using the method of Runge-Kutta. The obtained functions and the energy $E_0$, which contain the ground state and the kinetic energy of the transverse motion, are then used to calculate SOE with the use of formula
\ \\
\begin{equation}
|\Delta| = k_{\perp}\frac{\hbar^2}{2}<\Psi(z)|\left(\frac{\partial}{\partial z}\frac{1}{m(z)}\right)|\Psi(z)>
\end{equation}
\ \\
where $m(z)$ is given in Eq. (4). Since the potential is nonzero only at the interfaces and changes in the step-like fashion, the integration across the interfaces gives
\ \\
$$
|\Delta| = k_{\perp}\frac{\hbar^2}{2}\left\{|\Psi|^2\left(\frac{-a}{2}\right)\left[\frac{1}{m|_{{-a/2}^+}} - \frac{1}{m|_{{-a/2}^-}}\right] + \right.
$$
\ \\
\begin{equation}
\left. |\Psi|^2\left(\frac{a}{2}\right)\left[\frac{1}{m|_{{a/2}^+}} - \frac{1}{m|_{{a/2}^-}}\right]\right\}
\end{equation}
\ \\
 where $m|_{{-a/2}^-}=m_0[1+((E_0-V^l_b)/2m_0c^2]$; $m|_{{-a/2}^+}=m|_{{a/2}^-}=m_0[1+E_0/2m_0c^2]$, and $m|_{{a/2}^+}=m_0[1+(E_0-V^r_b)/2m_0c^2]$.
Here $E_0$ denotes the energy of the bound state and the offsets $V^l_b$ and $V^r_b$ are defined in Fig. 4. The values $|\Psi|^2(-a/2)$ and $|\Psi|^2(a/2)$ are not equal, it is seen that the nonvanishing $|\Delta|$ results also from the fact that the electron mass $m(z)$ depends on the potential $V(z)$ which is different at the left and right interface. This can be traced back to the above mentioned existence of the "mass force" appearing in special relativity. Thus both SIA and the relativistic effects are necessary to obtain nonvanishing spin-orbit energy for the bound states. As in the case of nonquantized spectrum, SOE for a one-dimensional potential $V(z)$ is proportional to the transverse wave vector $k_{\perp} = \sqrt{k_x^2 + k_y^2}$. It follows from Eq. (29) that the spin-orbit energy $\Delta$ is larger for more asymmetric wells, i. e. for larger difference between $V^l_b$ and $V^r_b$.

 We carry numerical estimations for $V^r_b = 5\cdot10^5$ eV, $V^l_b = 10^4$ eV and the width $a = 1\rm{\AA}$. Calculated energies for $k_{\perp} = k_x = 4.7\cdot10^8$ cm$^{-1}$ are $E_0 = 120$ eV and $\Delta$ = 0.7 meV. For a narrower well of the width $a$ = 0.1$\rm{\AA}$ and $k_{\perp} = k_x = 3.9\cdot10^9$ cm$^{-1}$ we calculate $E_0 = 8230$ eV and $\Delta$ = 3.2 eV.
\ \\
\section{\label{sec:level4}Discussion\protect\\ \lowercase{}}
\ \\

First, we briefly mention approximate features of our treatment. There is a certain ambiguity concerning the considered barrier. Most of the time the barrier is assumed to be vertical but, when estimating the spin-orbit energy, we assume its finite slope. This ambiguity is not troublesome because, in order to obtain measurable spin-orbit energy and perceptible difference of reflection angles (see Fig. 2), one needs a very high electric field, i.e. an \emph{almost vertical} barrier. One should also bear in mind that the employed procedure of treating separately the diagonal parts of the Hamiltonian and then calculating the matrix elements of the SOI nondiagonal parts is approximate. One should in principle solve exactly the two coupled differential equations, given by Eq. (3), applying the corresponding boundary conditions involving spin. However, the approximate procedure is known to give good results if the nondiagonal terms are small.

It was mentioned above, but we want to repeat it explicitly, that the reason for the spin splitting of energy due to the SOI in both considered cases is SIA of the systems. It is well known that if a system is characterized by both time and spatial inversion symmetries, the one-electron energies have at least double degeneracy. Both systems considered above possess the time reversal symmetry, it is SIA that causes the spin splitting via the spin-orbit interaction. However, in both cases the SOI is manifested only if there is also nonvanishing motion in the transverse direction characterized by $k_x$ and $k_y$.

We stress again that in the presence of spin-orbit interaction one deals with spin-mixed states characterized by effective spins (this was also remarked in Ref. 4). Thus, the spin-conserving and spin-flip reflections should be understood in terms of effective spins. Such processes are well known in semiconductor physics, where one deals with spin-mixed states due to the SOI and spin-flip scattering processes caused by electric perturbations like phonons, impurities, photons etc., see Ref. 12. The possibility of spin-flip processes due to the SOI bears the name of Elliott-Yafet mechanism. The special feature of the case we consider is that both the spin splitting of energies and the spin-flip reflection processes are caused by the same electric potential of the barrier via the spin-orbit interaction.

Observable effects of the SOI in both above cases occur for very small widths of the barrier and quantum well resulting in very high electric fields. One should bear in mind that in the hydrogen atom one deals with the binding energy of 13.6 eV corresponding to the atomic radius of around 0.5 $\rm{\AA}$ which corresponds to electric fields of about 3$\cdot10^9$ V/cm. In addition, quantum energies are smaller for 1D potentials of our interest here than for 3D potentials. Clearly, we chose energies and potentials smaller than $2m_0 c^2$ in order to avoid the effects of Klein paradox. It should be emphasized that, in both our cases, the proposed effects occur only for the relativistic electrons. First, because the spin-orbit interaction has the relativistic origin, second, because in an asymmetric quantum well the spin splitting due to the SOI is realized via the relativistic dependence of electron mass on external potential.

We mentioned above that the proposed system can serve as a source or a filter of spin-polarized electron beams since it spatially separates electrons with oriented effective spins. One can also say that this arrangement realizes the Stern-Gerlach experiment for free electrons in a vacuum which, in its original formulation with a magnetic field, remains a controversial problem [13, 14].

\section{\label{sec:level4}SUMMARY\protect\\ \lowercase{}}
\ \\

We described oblique reflection of spin-polarized relativistic electrons from a one-dimensional potential barrier taking into account the spin-orbit interaction existing in the Dirac equation. It is shown that the spin-conserving and spin-flip reflections have different reflection angles and they can serve as spin filters or spin polarizers. Numerical estimations of reflection angles and amplitudes are given. A one-dimensional asymmetric quantum well is considered and it is demonstrated that the energy of the electron bound state is split by the spin-orbit interaction if one takes into account relativistic dependence of electron mass on the energy and potential. General properties of energy splitting due to spin-orbit interaction are considered.
\ \\

{\bf Acknowledgments}

We are grateful to Dr T. M. Rusin for elucidating discussions.
\ \\

\appendix*
\section{\label{sec:level1}"Almost nonrelativistic" approximation \lowercase{}}

In order to
show in the simplest manner the physics of electron reflections from a barrier, we consider an "almost nonrelativistic" approximation in which the Schrodinger equation
is supplemented only by the spin-orbit interaction. The other two contributions appearing in the $v^2/c^2$ expansion, i.e. the Darwin and $p^4$ terms, are omitted because they are not important for our purposes.
In the above approximation the initial $2 \times 2$ eigenvalue equation for the positive electron energies reads
\ \\
\begin{equation}
\left\{\frac{{\bf{\hat{p}}}^2}{2m_0} + V({\bf{r}}) + \frac{\hbar}{4m_0^2c^2}\bm{\nabla} V\cdot
({\bf{\hat{p}}} \times {\bm{\hat{\sigma}}})\right\} \Phi = E \Phi\;\;,
\end{equation}
\ \\
in the standard notation, $m_0$ denotes the rest electron mass. The energy $E$ does not contain the rest electron energy.

Specifying the barrier we assume that the one-dimensional potential $V(z) = 0$ for $z \le 0$ and $V(z) = V_b$ for $z > 0$, where $V_b$ is barrier's height. When estimating the spin-orbit energy we somewhat modify this idealized picture, see Fig. 3. The motion in the $x$ and $y$ directions is free, so that $p_x \rightarrow \hbar k_x$ and $p_y \rightarrow \hbar k_y$. We consider an electron coming to the barrier from an oblique direction.

The eigenenergy equation (A.1) becomes
\ \\
$$
\left\{\left[-\frac{\hbar^2}{2m_0}\nabla^2 + V(z) - E\right]\left[\begin{array}{cc}
1&0 \\
0&1 \\ \end {array}\right]+\right.
$$
\ \\
\begin{equation}
\left. \frac{\hbar^2}{4m_0^2c^2}\frac{\partial V(z)}{\partial z}
\left[\begin{array}{cc}
0&-ik_x-k_y \\
ik_x-k_y&0\\ \end {array}\right]
\right\}\left|\begin{array}{c} a\Phi_{\uparrow} \\ b\Psi_{\downarrow}\end{array}\right| = 0
\;\;,
\end{equation}
where $a$ and $b$ are coefficients to be
determined, and $\Phi_{\uparrow}$ and $\Phi_{\downarrow}$ are the spin-up and spin-down states, as given in Eq. (5).
Without loss of a generality one can set $k_x \ne 0, k_y = 0$.

The transition from the pure spin states $\Phi_{\uparrow}$, $\Phi_{\downarrow}$ to the effective spin states $\Phi^1$, $\Phi^2$ is discussed in Eqs. (5)-(8). Suppose that the electron is initially in the effective spin-up state $\Phi^1$, so that its energy is
\ \\
\begin{equation}
E_1 = \frac{\hbar^2 k^2}{2m_0} + \Delta
\;\;,
\end{equation}
\ \\
where $k^2 = k_x^2 + k_z^2$. After an elastic reflection from the barrier there are two possibilities. If the reflected electron is still in the $\Phi^1$ state (spin-conserving process) its energy is again given by Eq. (A.3) and the resulting $k_x$ remains the same, while $k_z$ changes sign. This means that, in a spin-conserving reflection, the outgoing and incoming electron directions form the same angles with the normal to the barrier. If, on the other hand, the reflected electron is in the $\Phi^2$ state (spin-flip process), its energy is
\ \\
\begin{equation}
E_2 = \frac{\hbar^2 k'^2}{2m_0} - \Delta\;\;,
\end{equation}
\ \\
where ${k'}^2 = k_x^2 + {k'_z}^2$. In an elastic reflection the total electron energy must remain the same, so we have from Eqs. (A.3) and (A.4)
\ \\
\begin{equation}
\frac{\hbar^2}{2m_0}(k'^2 - k^2) = 2\Delta\;\;.
\end{equation}
\ \\
Since $k'_x = k_x$ because there is no force along the barrier (i.e. in the $x$ direction), we finally have
\ \\
\begin{equation}
{k'_z}^2 - k_z^2 = \frac{4m_0 \Delta}{\hbar^2}\;\;.
\end{equation}
\ \\
This means that, because of the SOI, in a spin-flip reflection the outgoing and incoming directions do \emph{not} form the same angles with the normal to the barrier. The above reasoning is illustrated in Fig. 1.
In addition, there exist two decaying waves penetrating the barrier: the effective spin-up wave with imaginary wave vector $q_z$
\ \\
\begin{equation}
q^2_z=(E - \Delta - V_b - \frac{\hbar^2 k^2_x}{2m_0})\frac{2m_0}{\hbar^2}\;\;,
\end{equation}
\ \\
and the effective spin-down wave with the imaginary wave vector $q'_z$
\ \\
\begin{equation}
q'^2_z=(E + \Delta - V_b - \frac{\hbar^2 k^2_x}{2m_0})\frac{2m_0}{\hbar^2}\;\;\;.
\end{equation}
\ \\
The complete wave functions have the form given in Eqs. (14) and (15) and the considerations of boundary conditions follow those given above for the Dirac equation. For the initial effective spin-up state one finally obtains the amplitudes in the following form
\ \\
\begin{equation}
R=\frac{[(k_z-q_z)(k'_z+q'_z)-S^2]}{[(k_z+q_z)(k'_z+q'_z)+S^2]}\;\;\;,
\end{equation}
\ \\
\begin{equation}
R'=\frac{-2S k_z}{[(k_z+q_z)(k'_z+q'_z)+S^2]}\;\;\;,
\end{equation}
\ \\
\begin{equation}
T=\frac{2k_z(k'_z+q'_z)}{[(k_z+q_z)(k'_z+q'_z)+S^2]}
\;\;\;,
\end{equation}
\ \\
\begin{equation}
T'=R'
\end{equation}
\ \\
It can be checked that for the nonrelativistic limit $E_E \approx E_V \approx 2m_0 c^2$ the relativistic formulas of Eqs. (22) - (25) reduce to Eqs. (A.9) - (A.12). The important point is that $R'$ does not vanish which means that some of the reflected electrons flip their effective spin.

\end{document}